\documentstyle[11pt,fleqn]{article}
\def\ref{par\noindent\hangindent=6mm\hangafter=1}
\begin{document}
\baselineskip 8mm
\begin{center}
{{\em Vestnik Leningradskogo Universiteta, No. 10, pp. 22-28, 1971}}
\end{center}
\begin{center}
{\bf  Study of the one-dimensional Schroedinger equation generated
from the hypergeometric equation}

\bigskip

G.A. Natanzon \\

Received 24 October 1970; published October 1971 \\

{\scriptsize Note: Translated from Russian by H.C. Rosu (November 1998)}

\end{center}

\bigskip
\bigskip


{\bf Original English Summary.} - A systematic method of constructing
potentials, for which
the one-variable Schroedinger equation can be solved in terms of the
hypergeometric (HGM)
function, is presented. All the potentials, obtained by energy-independent
transformations of the HGM equation, are determined together with eigenvalues
and eigenfunctions.
A class of potentials derived from the confluent HGM equation is found by
means of a limit process.


\bigskip




To study theoretically the rotational and rotational-vibrational
spectra of diatomic molecules, one often uses one-dimensional model
potentials for which the solution can be expressed in terms of the
HGM functions \cite{1}.
The problem of the changes of variable in the HGM equation leading to
the one dimensional Schroedinger equation has been studied several times in
the past \cite{1,2,3}, but the quoted authors have considered only
the case of potentials given in explicit form \footnote{The existence of
potentials $U(x)$, not explicitly depending on $x$, has been first posed
in \cite{1} (see also \cite{4}).}. However, knowing the energy spectrum
and the wavefunctions of the equation may prove useful for many applied
problems whether or not the potential is given in explicit or implicit
form.

The general form of the transformations leading from the HGM equation to the
Schroedinger equation is determined by the requirement ($z^{'}=dz/dx$)
\begin{equation} \label{1}
(z^{'})^{2}I(z)+\frac{1}{2}\{z,x\} =k^2-2MU(x)~,
\end{equation}
where
\begin{equation} \label{2}
I(z)=\frac{(1-\lambda _{0}^{2})(1-z)+(1-\lambda _{1}^{2})z+(\mu ^2-1)z(1-z)}
{4z^2(1-z)^2}~;
\end{equation}
and $\{z,x\}$ is the Schwartzian derivative of $z(x)$ with respect to $x$
\begin{equation} \label{3}
\{z,x\}=\frac{z^{''}}{z^{'}}\Bigg[\frac{z^{'''}}{z^{''}}
-\frac{3}{2}\frac{z^{''}}{z^{'}}\Bigg]~;
\end{equation}
$U(x)$ is the potential function; $M$ is the reduced mass; $E=\frac{k^2}{2M}$
is the energy. When condition (1) is fulfilled the wavefunction $\Psi$ is
related to the HGM function as follows
\begin{equation} \label{4}
\Psi[z(x)]=(z^{'})^{-1/2}z^{\frac{\lambda _{0}+1}{2}}
(1-z)^{\frac{\lambda _{1}+1}{2}}F(\alpha,\beta,\gamma;z)~.
\end{equation}
Here
\begin{equation} \label{5}
\left\{\begin{array}{ll}
\lambda _{0}=\gamma-1,\\
\lambda _{1}=\alpha +\beta-\gamma,\\
\mu=\beta-\alpha.
\end{array}
\right.
\end{equation}
We shall assume that $z(x)$ and therefore also $\{z,x\}$ do not depend on
the energy $E$. In this case, comparing the lhs and rhs of Eq.(1) one
concludes that the parameters $\mu ^{2}$, $\lambda _{0}^{2}$
and $\lambda _{1}^{2}$ are linear in $k^{2}$
\begin{equation} \label{6}
\left\{\begin{array}{ll}1-\mu ^2=ak^{2}-f\\
1-\lambda _{p}^{2}=c_{p}k^{2}-h_{p} & p=0,1
\end{array}
\right.
\end{equation}
and therefore $z(x)$ fulfills the differential equation
$$
\frac{(z^{'})^2R(z)}{4z^2(1-z)^2}=1
\eqno(7)
$$
where
$$
R(z)=a(z-1)z+c_{0}(1-z)+c_{1}z
\eqno(8a)
$$
or in a slightly different form
$$
R(z)=az^2+b_{0}z+c_{0}=a(z-1)^{2}+b_{1}(z-1)+c_{1}~.
\eqno(8b)
$$
Introducing in $\{z,x\}$ the logarithmic derivatives of $z^{'}$ and $z^{''}$
that can be found using Eq.(7), one next gets the potential from Eq.(1) by
means of Eq.(6)
$$
2MU[z(x)]=\frac{fz(z-1)+h_{0}(1-z)+h_1z+1}{R}+
\Bigg[a+\frac{a+(c_1-c_0)(2z-1)}{z(z-1)}-\frac{5}{4}\frac{\Delta}{R}\Bigg]
\frac{z^2(1-z)^2}{R^2}
\eqno(9)
$$
where the discriminant $\Delta=b_{p}^{2}-4ac_{p}=(a-c_{1}-c_{0})^2-4c_1c_{0}$.

Thus, one will get a {\em semiparametric} (i.e., including the constant of
integration of the differential equation) family of potential curves, for
one can always choose in the role of the three parameters, the scale factor,
the origin of the coordinate $x$ and the origin of the energy $E$.

For the Schroedinger solution given by Eq.(4) the variable $z$ is considered
within the interval [0,1]. The corresponding interval for the variable $x$
will be studied shortly.

Since the case of the potential well of infinite depth is not of much interest,
we shall ask the three-term quadratic polynomial $R(z)$ to have no
zeros for $0<z<1$, that is we shall consider
$$
R(z)>0~.
\eqno(10)
$$
It follows from Eq.(7) that in the unit interval the function $z(x)$ is
monotone so that the corresponding transformation is single valued. Eq.(10)
implies \footnote{For $\Delta >0, \quad a>0$ the coefficients $b_0$ and
$b_1$ should be of the same sign, i.e., $b_0b_1=(c_1-c_0)^2-a^2>0$.}
$$
c_{p}\geq0
\eqno(11a)
$$
$$
a<(\sqrt{c_1}+\sqrt{c_0})^{2}~.
\eqno(11b)
$$
Integrating Eq.(7) gives

a) $\Delta \neq 0$
$$
\pm 2x=\sum _{p=0}^{1}(-1)^{p}\Bigg[
\sqrt{c_{p}}\ln |b_p-2\sqrt{c_p}t_p|-b_p\int\frac{dt_p}{t_p^2-a}\Bigg]
\eqno(12a)
$$
where $t_p=\frac{\sqrt{r}-\sqrt{c_p}}{z-p}$;

b) $\Delta =0$
$$
 \pm 2(x-x_0)=\sum_{p=0}^{1}(-1)^{p}\sqrt{c_p}\ln|z-p|~.
\eqno(12b)
$$
The transformation $z(x)$ can be obtained in explicit form only for a few
particular values of the parameters $a$ and $c_{p}$. One should notice the
case when the zeros of $R$ coincide with the singularities of the HGM
equation. The resulting potentials have been already
considered by various authors as follows

a) Poeschl-Teller potentials \cite{5} ($R=b_0z(1-z)$ or $R=b_0$):
$$
2Mb_0U(x)=\left\{\begin{array}{ll}
f+1-\frac{h_0+\frac{3}{4}}{\sin ^2(x/\sqrt{b_0})}-
\frac{h_1+\frac{3}{4}}{\cos ^2(x/\sqrt{b_0})} \qquad (13a)\\
h_{1}+1+\frac{h_0+\frac{3}{4}}{{\rm sh}^2(x/\sqrt{b_0})}-
\frac{f+\frac{3}{4}}{{\rm ch}^2(x/\sqrt{b_0})} \qquad (13b)
\end{array}
\right.
$$

b) Rosen-Morse potentials \cite{6}  ($R=c_0$)
$$
2Mc_{0}U(x)=\frac{h_0+h_1+2}{2}+\frac{h_0-h_1}{2}{\rm th}(x/\sqrt{c_0})-
\frac{1}{4}\frac{f}{{\rm ch}^2(x/\sqrt{c_0})}
\eqno(14)
$$

c) Manning-Rosen potentials \cite{7} ($R=az^2$)
$$
2MaU(x)=\frac{f+h_1+2}{2}+\frac{f-h_1}{2}{\rm cth}(x/\sqrt{a})-
\frac{1}{4}\frac{h_0}{{\rm sh}^2(x/\sqrt{a})}~.
\eqno(15)
$$
The potentials (13-15) have the following general properties: the isotopic
shift can be described by means of some changes of the parameters $f$, $h_0$
and $h_1$. This property does not extend to the whole family of potential
curves given by Eq.(9). The scale transformation of the coordinate $x$ leads
in this case to an equation with five singular points, 0, 1, $z_1$, $z_2$,
and $\infty$, where $z_1$ and $z_2$ are the zeros of $R$.

Let us find now the interval where $x$ is defined for $\Delta\neq 0$. For
that, we first notice that the integral entering Eq. (12a) has no
singularities, because from $t_p=\sqrt{a}$ or
$$
\sqrt{a+\frac{b_p}{z-p}+\frac{c_{p}}{(z-p)^2}}=\sqrt{a}+\frac{c_p}{z-p}
$$
$\Delta =0$ would follow. This is why, performing
$z\rightarrow p$ ($t_p\rightarrow b_p/2\sqrt{c_p}$) in Eq. (12), we get
the limits $\pm \infty$ in $x$ for $c_p\neq 0$; if however one of the
coefficients $c_p$
is nought then the corresponding choice of the coordinate origin for $x$
as well as of the sign in Eq.(12a) can be done such that
$$
0\leq x<\infty~.
$$
Finally, the variable $x$ has a finite definition interval whenever both
$c_0$ and $c_1$ are zero.

The case when only one of the $c_p$ coefficients, for example $c_0$, is
zero, is of special interest for the theory of two atom molecules, because
for $x\rightarrow 0$ ($z\rightarrow 0$) the function $2Mb_0U(x)$ goes to
infinity as $(h_0+\frac{3}{4})z^{-1}$. For this potential, contrary to the
Morse case, the Schroedinger equation is solved with regular boundary
conditions for $x=0$ (see \cite{8}).

If $h_0=\frac{3}{4}$, the potential $U(x)$ is finite at the origin and for
$x\geq 0$ ($0\leq z\leq 1$) reads
$$
2MaU[z(x)]=\frac{f(z-1)+h_1+\frac{3}{4}}{z-z_1}+1+
\frac{3}{4}\frac{(3z_1-1)(1-z_1)}{(z-z_1)^2}-\frac{5}{4}
\frac{z_1(1-z_1)^2}{(z-z_1)^3}
\eqno(16)
$$
For $x<0$ it is natural to a priori determine the potential Eq.(16) to be
symmetric $U(x)=U(-x)$. Notice that for
$$
-\frac{1}{4}-\frac{9}{4}\frac{(1-z_1)}{z_1}>(f+\frac{3}{4})(1-z_1)>-1
$$
the potential Eq.(16) has two symmetric minima separated by a small barrier.

Since the HGM solution which is irregular
at $z=1$ has a $(1-z)^{-\lambda _1}$
($\lambda _1>0$) behavior at that point, the integral
$$
\int _{0}^{\infty}\Psi ^{2}dx=\frac{1}{4}\int _{0}^{1}Rz^{\lambda _0-1}
(1-z)^{\lambda _1-1}[F(\alpha,\beta,\gamma;z)]^2dz
\eqno(17)
$$
is finite only if $\alpha$  (or $\beta$) is equal either to a negative integer
or zero: $\alpha=-n$. The eigenfunctions $\Psi _{n}^{\pm}$ are
determined by the conditions
$$
\frac{d\Psi _{n}^{+}}{dx}|_{x=0}=0
\eqno(18a)
$$
and
$$
\Psi _{n}^{-}(0)=0~,
\eqno(18b)
$$
leading to a finite integral Eq. (17). The energy spectrum is determined by
$$
p+\frac{1}{2}+\sqrt{h_1+1-c_1k_{p}^{2}}=\sqrt{f+1-ak_{p}^{2}}~,
\eqno(19)
$$
where $p=2n$ for even levels and $p=2n+1$ for the odd ones, respectively.

Let us clarify now under what conditions the potential Eq.(9) leads to a
discrete spectrum. We shall consider $c_1\neq0$. For a discrete spectrum to
occur, the function $(z^{'})^{-\frac{1}{2}}\Psi$ should go to $\infty$ when
$z\rightarrow 0$ at least as $z^{-\frac{1}{2}}$, if one takes $\Psi$ as
the general solution of the Schroedinger equation. This condition is
obviously true for $c_0\neq0$. If $c_0=0$ the existence of the discrete
spectrum is possible only for positive values of $h_0$. The energy levels
come from the equation
$$
2n+1=\sqrt{f+1-ak_{n}^{2}}-\sqrt{h_0+1-c_0k_{n}^{2}}-
\sqrt{h_1+1-c_1k_{n}^{2}}~.
\eqno(20)
$$
It is easy to see that the spectrum is bounded from above, if and only if
both $c_0$ and $c_1$ are not zero, and therefore, it follows from Eq.(20) that
the discrete part of the spectrum has only a finite set of energy levels.
Only the potential curve Eq.(13a), for which $c_0=c_1=0$,
has an infinity of discrete levels.

The eigenfunctions of the discrete spectrum are of the form
$$
\Psi _n=B_nz^{\frac{\lambda _0-1}{2}}(1-z)^{\frac{\lambda _1-1}{2}}
(Rz^{'})^{\frac{1}{2}}P_{n}^{(\lambda _1,\lambda _2)}(2z-1)~.
\eqno(21)
$$
Here $P_{n}^{(\lambda _1,\lambda _0)}(2z-1)$ are the Jacobi polynomials,
whereas $B_n$ is a normalization constant given by
$$
B_{n}=\Bigg[\left(\frac{c_1}{\lambda_1}+\frac{c_0}{\lambda _0}
-\frac{a}{\mu}\right)\frac{\Gamma(\lambda _0+n+1)\Gamma(\lambda _1+n+1)}{
n!\Gamma(\mu-n)}\Bigg]^{-\frac{1}{2}}~,
\eqno(22)
$$
(see the integrals 7.391 (1) and (5) in \cite{9}). In particular,
the eigenfunctions $\Psi _{n}^{(\pm)}$, corresponding to the potential
Eq.(16), are obtained from Eq. (21)
for $c_0=0, \lambda _{0}=\pm \frac{1}{2}$.

We notice that the potential curve Eq.(13a) displays one more important
property: the Jacobi polynomials for neighbour energy levels are connected
by simple recurrence relationships (the parameters $\lambda _0$ and
$\lambda _1$ does not depend in this case on energy), and the matrix
elements $\langle m|z|n\rangle$ is different from zero for $m=n, n\pm1$.

To pass to the limit of the confluent HGM equation we make the following
notations
$$
a=\frac{\sigma _2}{\tau ^2},\quad c_1=\frac{\sigma _{2}}{\tau ^2}+
\frac{\sigma _1}{\tau}+c_0 \qquad (b_0=\frac{\sigma _1}{\tau}),
\eqno(23a)
$$
$$
f=\frac{g_2}{\tau ^2}, \quad h_1=\frac{g_2}{\tau ^2}+\frac{g_1}{\tau}+h_0
\eqno(23b)
$$
and
$$
z=\tau\zeta ~.
\eqno(24)
$$
For $\tau \rightarrow 0$ we have
$$
\frac{(\zeta ^{'})^2R(\zeta)}{4\zeta ^2}=1~,
\eqno(25)
$$
$$
I(\zeta)=-\frac{\delta _2}{4}-\frac{\delta _1}{4\zeta}
+\frac{1-\lambda _{0}^{2}}{4\zeta ^2}~,
\eqno(26)
$$
where
$$
\delta _1={\rm lim} _{\tau \rightarrow 0}[\tau(\lambda _{1}^{2}-\mu ^2)]=
g_{1}-\sigma _{1}k^2~,
\eqno(27a)
$$
$$
\delta _2={\rm lim}_{\tau \rightarrow 0}[\tau ^2\mu ^2]
=g_2-\sigma _2k^2\geq 0~.
\eqno(27b)
$$

For $\delta _2>0$ \footnote{For $\delta _2=0$, $\Psi=(\zeta ^{'})^{-1/2}
\zeta ^{1/2}Z_{\frac{\lambda _0}{2}}(\sqrt{-\delta _1}\zeta ^{1/2})$
(see the formula 2, 162 (1a) in \cite{10})} the wavefunction $\Psi$ is
connected with the Whittaker and the confluent series as follows
$$
\Psi=(\zeta ^{'})^{-1/2}M_{-\frac{\delta _1}{4\sqrt{\delta _0}},
\frac{\lambda _0}{2}}(\sqrt{\delta _2\zeta})
\eqno(28a)
$$
or
$$
\Psi=(\zeta ^{'})^{-1/2}(\sqrt{\delta _2}\zeta)^{\frac{\lambda _0+1}{2}}
e^{-\sqrt{\delta _2}\zeta/2}F(\frac{\gamma}{2}+\frac{\delta _{1}}
{4\sqrt{\delta _2}},\gamma;\sqrt{\delta _2\zeta})~.
\eqno(28b)
$$
The series terminates if
$$
-\frac{\gamma}{2}-\frac{\delta _1}{4\sqrt{\delta _2}}=n,\qquad n=0,1,2...
\eqno(29)
$$
The potential $U[\zeta(x)]$ takes the form
$$
2MU[\zeta(x)]=\frac{g_2\zeta^2+g_1\zeta+h_{0}+1}{R}+
\Bigg[\frac{\sigma _1}{\zeta}-\sigma _2 -\frac{5\Delta}{4R}\Bigg]
\frac{\zeta ^2}{R^2}~,
\eqno(30)
$$
where $\Delta=\sigma _{1}^{2}-4\sigma _2c_0$; here the function $\zeta (x)$
is given in the following implicit form
$$
\pm2(x-x_0)=\sqrt{R}+\sqrt{c_0}\ln |\sigma _1-2t\sqrt{c_0}|+
\frac{\sigma _1}{2\sqrt{\sigma _2}}\ln|\frac{t+\sqrt{\sigma _2}}
{t-\sqrt{\sigma _2}}|~,
\eqno(31)
$$
where $t=\frac{\sqrt{R}-\sqrt{c_0}}{\zeta}$.

It is supposed that $R(\zeta)$ has no zeros within $(0,\infty)$, that is
$$
\sigma _2\geq0,\qquad c_0\geq0, \qquad \sigma _{1}>-2\sqrt{c_0\sigma _2}~.
\eqno(32)
$$
If two of the three parameters $\sigma _1$, $\sigma _2$ and $c_0$ turn to
zero, Eq.(31) is easy to solve with respect to $\zeta$ and the potential leads
in this case to: a) a spherically symmetric oscillator well (or barrier)
for $\sigma _2=c_0=0$ \cite{11}; b) a Morse potential
for $\sigma _1=\sigma _2=0$ \cite{12}; a Kratzer potential for
$\sigma _1=c_0=0$ \cite{13,14}.

We remark that in the work \cite{15} an implicit potential occurs for
which the Schroedinger equation is turned into a confluent HGM equation.
This potential can be obtained from Eq.(30), if $\sigma _2=0$. The
assumptions in \cite{15} may be considered equivalent to a transformation
$\zeta(x)$ not depending on energy and therefore the results of \cite{15}
can be easily obtained in our framework.

Let us study now the definition interval of $x$. For $\zeta\rightarrow \infty$
$x$ goes to $\pm\infty$ as $\pm \frac{1}{2}\sqrt{\sigma _2}\zeta$, if
$\sigma _2\neq 0$, and as $\pm \sqrt{\sigma _1\zeta}$, if $\sigma _2=0$,
$\sigma _1\neq0$, because in the latter case
$$
\pm 2(x-x_0)=\sqrt{R}+\frac{\sigma _1}{t}+\sqrt{c_0}\ln|\zeta t^2|
\eqno(33)
$$
(for $\sigma _2=0$, $\sigma _1=0$ $x$ varies from $-\infty$ to $+\infty$).
For $\zeta \rightarrow 0$, $x\rightarrow \mp  \infty$ if $c_0\neq 0$, and
$x\rightarrow x_0$ if $c_0=0$.

For $\sigma _2\neq 0$, the zero of the energy is determined by the condition
$U(\infty)=0$, i.e., $g_2=0$. In this case, one gets for the energy spectrum
the following equation
$$
2n+1=-\frac{\sigma _1}{2\sqrt{\sigma _2}}|k_n|-\frac{g_1}{2\sqrt{\sigma _2}
|k_n|}-\sqrt{h_0+1-c_0k_{n}^{2}}~.
\eqno(34)
$$
From this, it follows that

a) for $g_1\geq0$, $h_0+1\geq0$, there is no discrete spectrum, because
$$
-\frac{\sigma _1}{2\sqrt{\sigma _2}}|k_n|-\sqrt{h_0+1-c_0k_{n}^{2}}\leq
\Bigg[\frac{-\sigma _1}{2\sqrt{\sigma _{2}}}-\sqrt{c_0}\Bigg]|k_{n}|\leq 0~;
\eqno(35)
$$

b) for $g_1\geq0$, $h _0+1<0$, a finite number of discrete levels is possible
if $\sigma _1<0$;

c) for $g_1<0$ the discrete spectrum has an infinite number of levels
converging to zero. For $\sigma _2=0$ the discrete levels are determined
by the equation
$$
1+h_0-c_0k_{n}^{2}=\left(2n+1+\frac{g_1-\sigma _1k_{n}^{2}}{2\sqrt{g_2}}
\right)^2
\eqno(36)
$$
and for $c_0\neq 0$ the discrete spectrum may have only a finite number of
energy levels (the case $\sigma _2=0$, $c_{0}=0$ corresponds to an infinite
number of levels).

The wavefunctions of the discrete spectrum are given in terms of the Laguerre
polynomials as follows
$$
\Psi _{n}=B_{n}R^{\frac{1}{2}}(\sqrt{\delta _2}\zeta)^{\frac{\lambda _0-1}{2}}
e^{-\sqrt{\delta _2}\frac{\zeta}{2}}(\sqrt{\delta _2}\zeta ^{'})^{\frac{1}{2}}
L_{n}^{(\lambda _0)}(\sqrt{\delta _2}\zeta)~.
\eqno(37)
$$
Here, the normalization factor $B_n$ reads
$$
B_n=\Bigg[\left(\frac{c_0}{\lambda _0}+\frac{\sigma _1}{\sqrt{\delta _2}}+
(\lambda _0+2n+1)\frac{\sigma _2}{\delta _2}\right)\frac{\Gamma(\gamma +n)}
{n!}\Bigg]^{-\frac{1}{2}}~.
\eqno(38)
$$

The integrals required for its calculation can be obtained from $I_{\nu}$
\cite{16}:
$$
I_{\nu}=\int e^{-\sqrt{\delta _2}\zeta}\zeta ^{\nu -1}[F(-n,\gamma;
\sqrt{\delta _2}\zeta)]^2d\zeta
\eqno(39)
$$
for $\nu=\gamma\pm1,\gamma$.

The potentials given by Eqs. (9) and (30) may have useful applications in
the theory of diatomic molecules, in particular, for the calculations of
important quantities such as the Franck-Condon factors and anharmonic
constants.

\noindent
{\bf Acknowledgments}

\noindent
The author is deeply grateful to M.N. Adamov, Yu.N. Demkov, and I.V. Komarov
for their critical remarks.


\end{document}